\documentclass{aastex63}

\usepackage[utf8x]{inputenc}
\usepackage{amssymb}
\usepackage{amsmath}
\usepackage{natbib}
\usepackage{hyperref}

\received{\today}
\revised{\today}
\accepted{\today}
\submitjournal{ApJL}

\shorttitle{First SO -- PSP Quadrature}
\shortauthors{Telloni et al.}

\begin{document}

\title{Exploring the Solar Wind from its Source on the Corona into the Inner Heliosphere\\
during the First Solar Orbiter -- Parker Solar Probe Quadrature}

\correspondingauthor{Daniele Telloni}
\email{daniele.telloni@inaf.it}

\author[0000-0002-6710-8142]{Daniele Telloni}
\affil{National Institute for Astrophysics, Astrophysical Observatory of Torino, Via Osservatorio 20, I-10025 Pino Torinese, Italy}
\author[0000-0003-1962-9741]{Vincenzo Andretta}
\affil{National Institute for Astrophysics, Astronomical Observatory of Capodimonte, Salita Moiariello 16, I-80131 Napoli, Italy}
\author[0000-0003-4155-6542]{Ester Antonucci}
\affil{National Institute for Astrophysics, Astrophysical Observatory of Torino, Via Osservatorio 20, I-10025 Pino Torinese, Italy}
\author[0000-0001-5796-5653]{Alessandro Bemporad}
\affil{National Institute for Astrophysics, Astrophysical Observatory of Torino, Via Osservatorio 20, I-10025 Pino Torinese, Italy}
\author[0000-0002-8430-8218]{Giuseppe E. Capuano}
\affil{National Institute for Astrophysics, Astrophysical Observatory of Catania, Via Santa Sofia 78, I-95123 Catania, Italy}
\affil{University of Catania, Department of Physics and Astronomy, Via Santa Sofia 64, I-95123 Catania, Italy}
\author[0000-0002-2789-816X]{Silvano Fineschi}
\affil{National Institute for Astrophysics, Astrophysical Observatory of Torino, Via Osservatorio 20, I-10025 Pino Torinese, Italy}
\author[0000-0002-3468-8566]{Silvio Giordano}
\affil{National Institute for Astrophysics, Astrophysical Observatory of Torino, Via Osservatorio 20, I-10025 Pino Torinese, Italy}
\author[0000-0003-4089-9316]{Shadia Habbal}
\affil{Institute for Astronomy, University of Hawaii, Honolulu, HI 96822, USA}
\author[0000-0003-1059-4853]{Denise Perrone}
\affil{Italian Space Agency, Via del Politecnico snc, I-00133 Roma, Italy}
\author[0000-0001-8247-7168]{Rui F. Pinto}
\affil{D\'epartement d’Astrophysique, Astrophysique Instrumentation Mod\'elisation, Commissariat \`a l'\'Energie Atomique et aux \'Energies Alternatives, Institut de Recherche sur les lois Fondamentales de l'Univers, Centre National de la Recherche Scientifique, Institut National des Sciences de l'Univers, Universit\'e Paris-Saclay, Universit\'e de Paris, Orme des Merisiers B\^at 709, F-91191 Gif-sur-Yvette, France}
\affil{Institut de Recherche en Astrophysique et Plan\'etologie, Universit\'e de Toulouse, Universit\'e fe Paul Sabatier, Midi-Pyr\'en\'ees Observatory, Centre National de la Recherche Scientifique, 9 Avenue du Colonel Roche, F-31028 Toulouse, France}
\author[0000-0002-5981-7758]{Luca Sorriso-Valvo}
\affil{Swedish Institute of Space Physics, \AA ngstr\"om Laboratory, L\"agerhyddsv\"agen 1, SE-751 21 Uppsala, Sweden}
\affil{National Research Council, Institute for the Science and Technology of Plasmas, Via Amendola 122/D, I-70126 Bari, Italy}
\author[0000-0003-3517-8688]{Daniele Spadaro}
\affil{National Institute for Astrophysics, Astrophysical Observatory of Catania, Via Santa Sofia 78, I-95123 Catania, Italy}
\author[0000-0002-1017-7163]{Roberto Susino}
\affil{National Institute for Astrophysics, Astrophysical Observatory of Torino, Via Osservatorio 20, I-10025 Pino Torinese, Italy}
\author[0000-0003-2845-4250]{Lloyd D. Woodham}
\affil{Department of Physics, Imperial College London, London SW7 2AZ, UK}
\author[0000-0002-4642-6192]{Gary P. Zank}
\affil{Center for Space Plasma and Aeronomic Research, University of Alabama in Huntsville, Huntsville, AL 35805, USA}
\affil{Department of Space Science, University of Alabama in Huntsville, Huntsville, AL 35805, USA}
\author[0000-0001-9921-1198]{Marco Romoli}
\affil{University of Florence, Department of Physics and Astronomy, Via Giovanni Sansone 1, I-50019 Sesto Fiorentino, Italy}
\author[0000-0002-1989-3596]{Stuart D. Bale}
\affil{Space Sciences Laboratory, University of California, Berkeley, CA 94720, USA}
\affil{Physics Department, University of California, Berkeley, CA 94720, USA}
\author[0000-0002-7077-930X]{Justin C. Kasper}
\affil{BWX Technologies, Inc., Washington, DC 20002, USA}
\affil{Climate and Space Sciences and Engineering, University of Michigan, Ann Arbor, MI 48109, USA}
\author[0000-0003-0972-7022]{Fr\'ed\'eric Auch\`ere}
\affil{Universit\'e Paris-Saclay, Centre National de la Recherche Scientifique, Institut d’Astrophysique Spatiale, Rue Jean-Dominique Cassini, F-91440 Bures-sur-Yvette, France}
\author[0000-0002-2152-0115]{Roberto Bruno}
\affil{National Institute for Astrophysics, Institute for Space Astrophysics and Planetology, Via del Fosso del Cavaliere 100, I-00133 Roma, Italy}
\author[0000-0003-0520-2528]{Gerardo Capobianco}
\affil{National Institute for Astrophysics, Astrophysical Observatory of Torino, Via Osservatorio 20, I-10025 Pino Torinese, Italy}
\author[0000-0002-3520-4041]{Anthony W. Case}
\affil{Harvard-Smithsonian Center for Astrophysics, Cambridge, MA 02138, USA}
\author[0000-0001-8783-0047]{Chiara Casini}
\affil{National Research Council, Institute for Photonics and Nanotechnologies, Via Trasea 7, I-35131 Padova, Italy}
\affil{Centre of Studies and Activities for Space ``Giuseppe Colombo'', Via Venezia 15, I-35131 Padova, Italy}
\author[0000-0002-9716-3820]{Marta Casti}
\affil{The Catholic University of America at the National Aeronautics and Space Administration, Goddard Space Flight Center, Greenbelt, MD 20771, USA}
\author[0000-0002-3379-2142]{Paolo Chioetto}
\affil{National Research Council, Institute for Photonics and Nanotechnologies, Via Trasea 7, I-35131 Padova, Italy}
\author[0000-0003-0378-9249]{Alain J. Corso}
\affil{National Research Council, Institute for Photonics and Nanotechnologies, Via Trasea 7, I-35131 Padova, Italy}
\author[0000-0001-6273-8738]{Vania Da Deppo}
\affil{National Research Council, Institute for Photonics and Nanotechnologies, Via Trasea 7, I-35131 Padova, Italy}
\author[0000-0003-2426-2112]{Yara De Leo}
\affil{Max Planck Institute for Solar System Research, Justus-von-Liebig-Weg 3, D-37077 G\"ottingen, Germany}
\affil{University of Catania, Department of Physics and Astronomy, Via Santa Sofia 64, I-95123 Catania, Italy}
\author[0000-0002-4401-0943]{Thierry Dudok de Wit}
\affil{Laboratoire de Physique et de Chimie de l'Environnement et de l'Espace, Center National de la Recherche Scientifique, 3A Avenue de la Recherche Scientifique, F-45071 Orl\'eans, France}
\author[0000-0001-9014-614X]{Federica Frassati}
\affil{National Institute for Astrophysics, Astrophysical Observatory of Torino, Via Osservatorio 20, I-10025 Pino Torinese, Italy}
\author[0000-0001-5528-1995]{Fabio Frassetto}
\affil{National Research Council, Institute for Photonics and Nanotechnologies, Via Trasea 7, I-35131 Padova, Italy}
\author[0000-0003-0420-3633]{Keith Goetz}
\affil{School of Physics and Astronomy, University of Minnesota, Minneapolis, MN 55455, USA}
\author[0000-0002-1837-2262]{Salvo L. Guglielmino}
\affil{National Institute for Astrophysics, Astrophysical Observatory of Catania, Via Santa Sofia 78, I-95123 Catania, Italy}
\author[0000-0002-6938-0166]{Peter R. Harvey}
\affil{Space Sciences Laboratory, University of California, Berkeley, CA 94720, USA}
\author[0000-0002-5778-2600]{Petr Heinzel}
\affil{Czech Academy of Sciences, Astronomical Institute, Fri\v{c}ova 298, CZ-25165 Ond\v{r}ejov, Czechia}
\author[0000-0002-0764-7929]{Giovanna Jerse}
\affil{National Institute for Astrophysics, Astronomical Observatory of Trieste, Localit\`a Basovizza 302, I-34149 Trieste, Italy}
\author[0000-0001-6095-2490]{Kelly E. Korreck}
\affil{Harvard-Smithsonian Center for Astrophysics, Cambridge, MA 02138, USA}
\author[0000-0001-8244-9749]{Federico Landini}
\affil{National Institute for Astrophysics, Astrophysical Observatory of Torino, Via Osservatorio 20, I-10025 Pino Torinese, Italy}
\author[0000-0001-5030-6030]{Davin Larson}
\affil{Space Sciences Laboratory, University of California, Berkeley, CA 94720, USA}
\author[0000-0002-0016-7594]{Alessandro Liberatore}
\affil{National Institute for Astrophysics, Astrophysical Observatory of Torino, Via Osservatorio 20, I-10025 Pino Torinese, Italy}
\author[0000-0002-0396-0547]{Roberto Livi}
\affil{Space Sciences Laboratory, University of California, Berkeley, CA 94720, USA}
\author[0000-0003-3112-4201]{Robert J. MacDowall}
\affil{National Aeronautics and Space Administration, Goddard Space Flight Center, Greenbelt, MD 20771, USA}
\author[0000-0002-0901-0251]{Enrico Magli}
\affil{Politecnico of Turin, Department of Electronics and Telecommunications, Corso Duca degli Abruzzi 24, I-10129 Torino, Italy}
\author[0000-0003-1191-1558]{David M. Malaspina}
\affil{Astrophysical and Planetary Sciences Department, University of Colorado, Boulder, CO 80309, USA}
\affil{Laboratory for Atmospheric and Space Physics, University of Colorado, Boulder, CO 80303, USA}
\author[0000-0002-2656-1557]{Giuseppe Massone}
\affil{National Institute for Astrophysics, Astrophysical Observatory of Torino, Via Osservatorio 20, I-10025 Pino Torinese, Italy}
\author[0000-0002-5422-1963]{Mauro Messerotti}
\affil{National Institute for Astrophysics, Astronomical Observatory of Trieste, Localit\`a Basovizza 302, I-34149 Trieste, Italy}
\affil{University of Trieste, Department of Physics, Via Alfonso Valerio 2, I-34127 Trieste, Italy}
\author[0000-0001-9670-2063]{John D. Moses}
\affil{National Aeronautics and Space Administration, Headquarters, Washington, DC 20546, USA}
\author[0000-0003-2007-3138]{Giampiero Naletto}
\affil{University of Padua, Department of Physics and Astronomy, Via Francesco Marzolo 8, I-35131 Padova, Italy}
\author[0000-0002-9459-3841]{Gianalfredo Nicolini}
\affil{National Institute for Astrophysics, Astrophysical Observatory of Torino, Via Osservatorio 20, I-10025 Pino Torinese, Italy}
\author[0000-0003-2566-2820]{Giuseppe Nistic\`o}
\affil{University of Calabria, Department of Physics, Ponte Pietro Bucci Cubo 31C, I-87036 Rende, Italy}
\author[0000-0002-4440-7166]{Olga Panasenco}
\affil{Advanced Heliophysics, Pasadena, CA 91106, USA}
\author[0000-0002-3789-2482]{Maurizio Pancrazzi}
\affil{National Institute for Astrophysics, Astrophysical Observatory of Torino, Via Osservatorio 20, I-10025 Pino Torinese, Italy}
\author[0000-0002-1383-6750]{Maria G. Pelizzo}
\affil{National Research Council, Institute of Electronics, Information Engineering and Telecommunications, Via Gradenigo 6/B, I-35131 Padova, Italy}
\author[0000-0002-1573-7457]{Marc Pulupa}
\affil{Space Sciences Laboratory, University of California, Berkeley, CA 94720, USA}
\author[0000-0002-1820-4824]{Fabio Reale}
\affil{University of Palermo, Department of Physics and Chemistry - Emilio Segr\`e, Piazza del Parlamento 1, I-90134 Palermo, Italy}
\affil{National Institute for Astrophysics, Astronomical Observatory of Palermo, Piazza del Parlamento 1, I-90134 Palermo, Italy}
\author[0000-0001-7066-6674]{Paolo Romano}
\affil{National Institute for Astrophysics, Astrophysical Observatory of Catania, Via Santa Sofia 78, I-95123 Catania, Italy}
\author[0000-0002-5163-5837]{Clementina Sasso}
\affil{National Institute for Astrophysics, Astronomical Observatory of Capodimonte, Salita Moiariello 16, I-80131 Napoli, Italy}
\author[0000-0001-6060-9078]{Udo Sch\"uhle}
\affil{Max Planck Institute for Solar System Research, Justus-von-Liebig-Weg 3, D-37077 G\"ottingen, Germany}
\author[0000-0002-5365-7546]{Marco Stangalini}
\affil{Italian Space Agency, Via del Politecnico snc, I-00133 Roma, Italy}
\author[0000-0002-7728-0085]{Michael L. Stevens}
\affil{Harvard-Smithsonian Center for Astrophysics, Cambridge, MA 02138, USA}
\author[0000-0002-5425-7122]{Leonard Strachan}
\affil{Naval Research Laboratory, Space Science Division, Washington, DC 20375, USA}
\author[0000-0002-6280-806X]{Thomas Straus}
\affil{National Institute for Astrophysics, Astronomical Observatory of Capodimonte, Salita Moiariello 16, I-80131 Napoli, Italy}
\author[0000-0001-7298-2320]{Luca Teriaca}
\affil{Max Planck Institute for Solar System Research, Justus-von-Liebig-Weg 3, D-37077 G\"ottingen, Germany}
\author[0000-0002-7585-8605]{Michela Uslenghi}
\affil{National Institute for Astrophysics, Institute of Space Astrophysics and Cosmic Physics of Milan, Via Alfonso Corti 12, I-20133 Milano, Italy}
\author[0000-0002-2381-3106]{Marco Velli}
\affil{Earth, Planetary, and Space Sciences, University of California, Los Angeles, CA 90095, USA}
\author[0000-0002-0497-1096]{Daniel Verscharen}
\affil{Mullard Space Science Laboratory, University College London, Holmbury St. Mary, RH5 6NT Dorking, UK}
\affil{Space Science Center, Institute for the Study of Earth, Oceans, and Space, University of New Hampshire, Durham, NH 03824, USA}
\author[0000-0002-4997-1460]{Cosimo A. Volpicelli}
\affil{National Institute for Astrophysics, Astrophysical Observatory of Torino, Via Osservatorio 20, I-10025 Pino Torinese, Italy}
\author[0000-0002-7287-5098]{Phyllis Whittlesey}
\affil{Space Sciences Laboratory, University of California, Berkeley, CA 94720, USA}
\author[0000-0002-4184-2031]{Luca Zangrilli}
\affil{National Institute for Astrophysics, Astrophysical Observatory of Torino, Via Osservatorio 20, I-10025 Pino Torinese, Italy}
\author[0000-0002-9207-2647]{Gaetano Zimbardo}
\affil{University of Calabria, Department of Physics, Ponte Pietro Bucci Cubo 31C, I-87036 Rende, Italy}
\author[0000-0003-0290-3193]{Paola Zuppella}
\affil{National Research Council, Institute for Photonics and Nanotechnologies, Via Trasea 7, I-35131 Padova, Italy}

\begin{abstract}
This Letter addresses the first Solar Orbiter (SO) -- Parker Solar Probe (PSP) quadrature, occurring on January $18$, $2021$, to investigate the evolution of solar wind from the extended corona to the inner heliosphere. Assuming ballistic propagation, the same plasma volume observed remotely in corona at altitudes between $3.5$ and $6.3$ solar radii above the solar limb with the Metis coronagraph on SO can be tracked to PSP, orbiting at $0.1$ au, thus allowing the local properties of the solar wind to be linked to the coronal source region from where it originated. Thanks to the close approach of PSP to the Sun and the simultaneous Metis observation of the solar corona, the flow-aligned magnetic field and the bulk kinetic energy flux density can be empirically inferred along the coronal current sheet with an unprecedented accuracy, allowing in particular estimation of the Alfv\'en radius at $8.7$ solar radii during the time of this event. This is thus the very first study of the same solar wind plasma as it expands from the sub-Alfv\'enic solar corona to just above the Alfv\'en surface.
\end{abstract}


\section{Introduction}
\label{sec:introduction}
Remote sensing observations are essential to study the global magnetic configuration of the solar corona and to explore the acceleration regions of the solar wind \citep{antonucci2020a}. However, complications in integrating along the Line Of Sight \citep[LOS, e.g.][]{dolei2018} over an extended region of the corona \citep[an optically thin plasma, e.g.][]{bradshaw2013} and the low cadence measurements of coronagraphs make it difficult to investigate in detail local coronal heating, identify high-frequency waves of different kinds, or study turbulent properties of the coronal plasma as a magnetohydrodynamic fluid. On the other hand, in-situ measurements provide direct physical parameters of the plasma at a single point in space and time, enabling the examination of the local mechanisms that control, e.g. energy dissipation in the solar wind \citep{bruno2013,verscharen2019}. However, single-spacecraft in-situ measurements only cannot provide topological information about the magnetic field, and thus a three dimensional view of the plasma surrounding the spacecraft. Therefore, the combination of remote sensing observations and in-situ measurements provides an unequaled opportunity to connect solar wind source regions to the local properties of plasma and electromagnetic fields. This link works at best for in-situ measurements close to the Sun, where solar plasma is not yet fully reprocessed by stream-stream interactions.

There are several possible orbital configurations between two or more spacecraft relative to the Sun that allow the study of the connection of the plasma observed in situ to its solar source. One interesting case is provided by ``quadratures'', i.e. when the angular separation of two spacecraft with the Sun as reference point is $90^{\circ}$. Important examples of this kind of quadrature are given by the SOlar and Heliospheric Observatory \citep[SOHO,][]{domingo1995} -- Sun -- Ulysses \citep{wenzel1992} campaigns, when the same plasma remotely observed with SOHO was sampled in situ when it left the coronal source in the direction of Ulysses \citep{suess2000,suess2001,poletto2002}. The comparison of plasma properties in corona and heliosphere was made possibile thanks to the combined use of the Large Angle Spectroscopic COronagraph \citep[LASCO,][]{brueckner1995} to understand the overall coronal configuration at the time of quadratures, and the UltraViolet Coronagraphic Spectrometer \citep[UVCS,][]{kohl1995} to derive its physical characteristics, on SOHO, and the Solar Wind Ion Composition Spectrometer \citep[SWICS,][]{gloeckler1992} to provide interplanetary data, on Ulysses. Multiple quadratures (twice a year) during the SOHO -- Ulysses era allowed the study of solar wind parameters, such as density, speed and temperature, and their evolution from the Sun to the heliosphere. Other quantities, such as ion abundances or the ionization state, which instead remain almost unchanged during solar wind expansion, have been used to conclusively link a plasma stream measured in situ with the coronal source as identified in coronagraphic images \citep{bemporad2013}.

The launch of Parker Solar Probe \citep[PSP,][]{fox2016} in August $2018$ and Solar Orbiter \citep[SO,][]{muller2020} in February $2020$, offers an exciting new era to explore the connection between the Sun and the heliosphere. On the one hand, PSP will be the first spacecraft to fly into the extended solar corona, a yet unexplored environment with in-situ experiments. On the other hand, SO is the first spacecraft to carry both remote sensing and in-situ instruments at distances as close as $0.28$ au. It will leave the ecliptic plane, allowing for the first time observation of the solar poles, the regions where the fast solar wind originates (during solar minimum) and where magnetic fields cyclically reverse \citep{zouganelis2020}. The importance of this new era for solar physics also lies in the synergy between these two missions \citep{velli2020}, which will allow the tracing of the magnetic connectivity of transient outflows and the continuous solar wind to their solar sources. Particular attention has been paid to radial alignments between the two probes. Indeed, the first PSP -- SO line-up occurred in September $2020$, allowing the radial evolution of solar wind turbulence between $0.1$ and $1$ au to be studied, i.e. from a highly Alfv\'enic and less developed turbulence state near the Sun to a fully developed and intermittent turbulence state close to the Earth \citep{telloni2021}.

As far as it concerns quadrature configurations between the two spacecraft, with SO remotely observing the source region of the solar wind that is measured in situ with PSP, the first SO -- Sun -- PSP quadrature occurred on January $18$, $2021$. At that time, when traveling along its orbit very close to the Sun, PSP crossed the very inner heliosphere at a heliocentric distance just above $20$ solar radii (R$_{\odot}$), while the coronagraph Metis \citep{antonucci2020b} on SO was observing the solar atmosphere in an annular Field Of View (FOV) extending from $3.5$ to $6.3$ R$_{\odot}$ above the solar limb. It is thus the first time that an expanding coronal plasma, observed remotely with Metis, is almost immediately measured as it propagates outward with a suite of in-situ instruments.

This Letter presents joint SO -- PSP observations that describe the evolution of the pristine solar wind, not yet reprocessed by nonlinear interactions, from the corona into the very inner heliosphere, aiming to link the local properties of the plasma stream measured in situ with PSP with the coronal source from where it originated imaged by the SO/Metis coronagraph.

\section{Analysis and results}
\label{sec:analysis_results}
\begin{figure}[h]
	\begin{center}
		\includegraphics[width=0.5\linewidth]{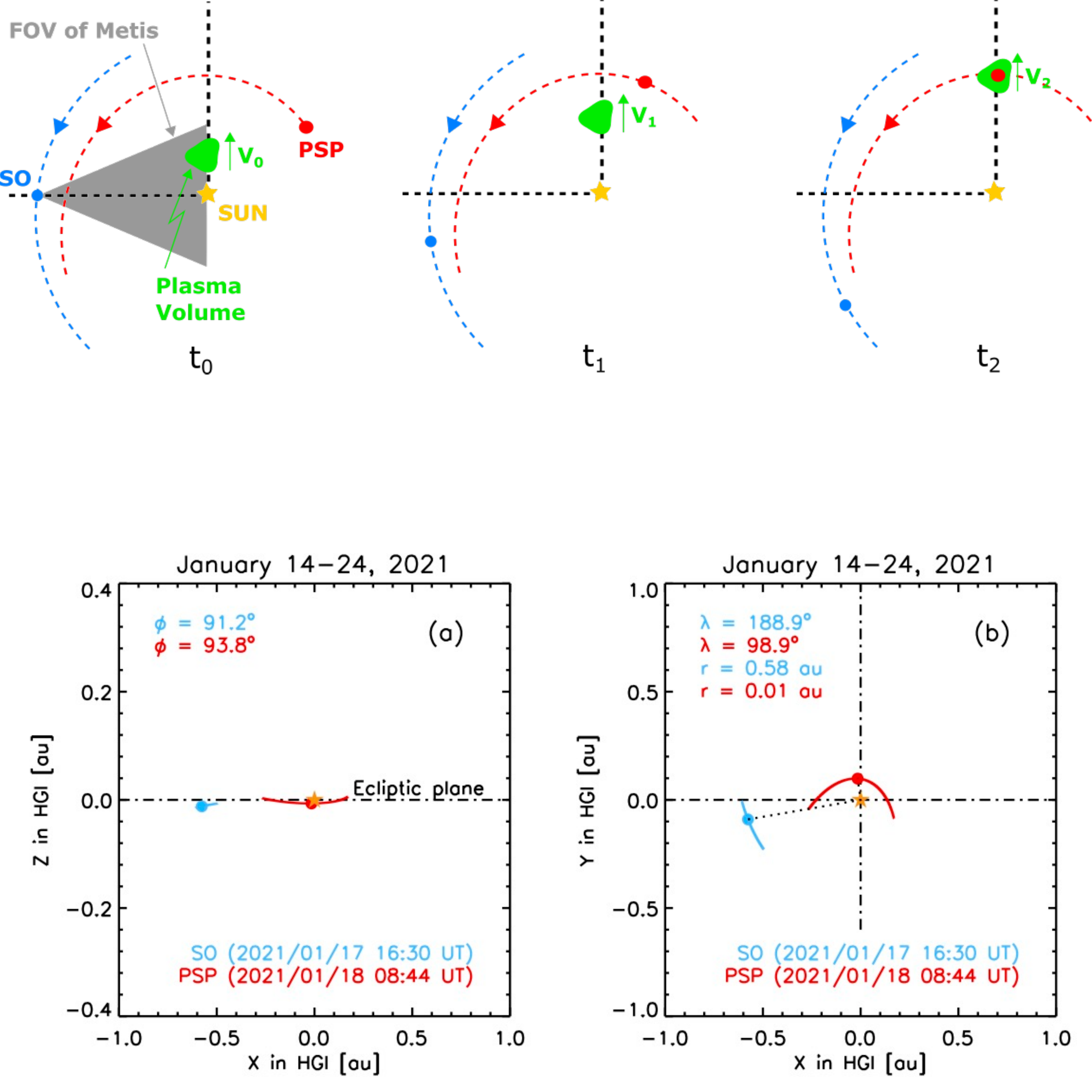}
	\end{center}
	\caption{Top panel: Cartoon showing how the same plasma volume observed remotely with SO/Metis can be measured locally with PSP during their quadrature. Bottom panels: SO (blue) and PSP (red) positions relative to the Sun (yellow star), in side (a) and top (b) views of the ecliptic plane in the HGI coordinate system, at the times (reported in the legends) of the observation of the same solar wind stream. The corresponding heliocentric distances ($r$), and heliographic latitudes ($\phi$) and longitudes ($\lambda$) are also reported. The spacecraft trajectories from January $14-21$, $2021$ and their longitudinal separation of $90^{\circ}$ in the XY plane are shown as color-coded solid and black dotted lines, respectively.}
	\label{fig:quadrature}
\end{figure}

The SO -- PSP orbital configuration offers the opportunity to track the same plasma volume as it expands from the extended corona to the inner heliosphere. Indeed, thanks to the continuous expansion of the solar corona, the two probes will at some point be located such that the plasma crossed by PSP, which is moving outward at a speed around $100-200$ km s$^{-1}$ on the solar equatorial plane, is the same plasma observed with Metis just a few hours earlier at a distance of $3.5-6.3$ R$_{\odot}$. From the cartoon of Fig. \ref{fig:quadrature} (top panel), this condition requires that during Metis observations, PSP is not yet on the instantaneous Plane Of the Sky (POS) of the coronagraph, so that PSP and the Metis-observed volume of plasma reach the same location at the same time.

On the basis of the spacecraft's ephemerides and (as shown below) of the solar wind speed at both coronal and heliospheric heights, this particular condition occurred when Metis was observing the solar corona on January $17$, $2021$ at $16:30$ UT ($t_{0}$ in the cartoon of Fig. \ref{fig:quadrature}). At that time PSP was about $30^{\circ}$ behind the instantaneous POS of Metis and it took $\sim16.2$ h to travel that longitudinal separation, thus reaching the Metis POS on January $18$, $2021$ at $08:44$ UT ($t_{2}$ in the cartoon of Fig. \ref{fig:quadrature}). This occurred when PSP was approaching perihelion at $0.1$ au from the Sun and at a latitude of $3.8^{\circ}$ below the ecliptic plane (i.e. $93.8^{\circ}$ counterclockwise from the North pole). As shown below (based on the expansion velocity inferred from Metis observations), the plasma observed (at $t_{0}$) with Metis at a distance of $3.5-6.3$ R$_{\odot}$ took $16.3\pm1.1$ h to travel the distance from PSP thus encountering it on its way outward. The corresponding SO and PSP locations are shown along their orbits in the XZ and XY planes of the HelioGraphic Inertial coordinate frame (HGI) of Fig. \ref{fig:quadrature}(a) and (b), respectively.

\begin{figure*}[h]
	\begin{center}
		\includegraphics[width=\linewidth]{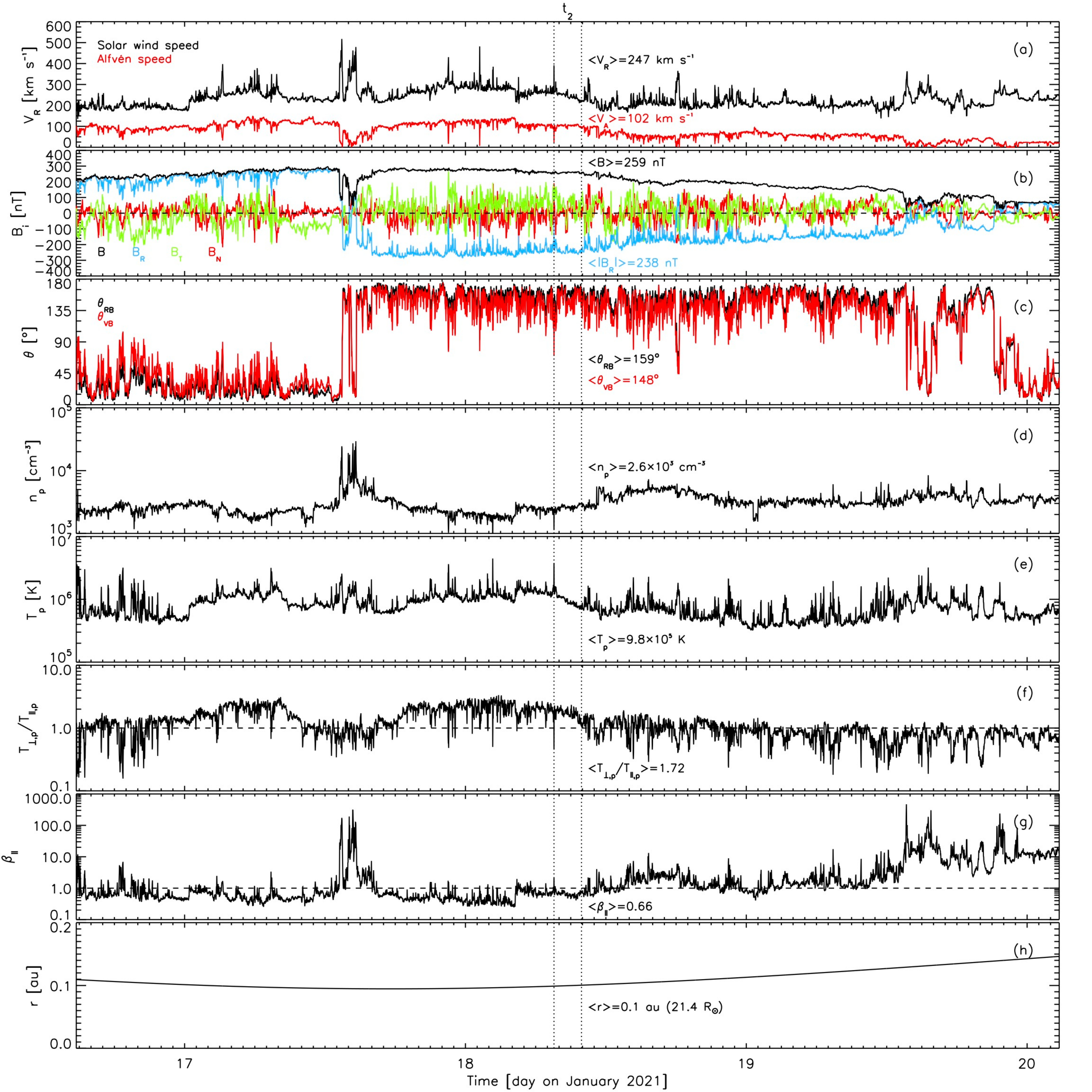}
	\end{center}
	\caption{PSP solar wind parameters over the $3.5$-day window centered on the $\sim2.5$ h long selected time interval (delimited by vertical dotted lines) around $t_{2}$: solar wind radial speed $V_{R}$ and Alfv\'en speed $V_{A}$ (a), magnetic field vector $\mathbf{B}$ (b), angles $\theta_{RB}$ and $\theta_{VB}$ of $\mathbf{B}$ with the radial and velocity directions (c), proton number density $n_{p}$ (d), temperature $T_{p}$ (e), temperature anisotropy $T_{\perp,p}/T_{\parallel,p}$ (f), and parallel plasma beta $\beta_{\parallel}$ (g), and radial distance $r$ (h). The average values reported within each panel refer to the identified period.}
	\label{fig:psp}
\end{figure*}

A time interval of about $2.5$ hours centered on $t_{2}$, comparable to the coronal plasma transit time through the Metis FOV (according to the speed of the outflowing plasma shown below) and resulting in a slight off-quadrature of $\pm2^{\circ}$ longitude relative to the Metis POS, is thus identified in PSP data as corresponding to the same plasma volume observed remotely with Metis $16.3$ h earlier. Some relevant parameters during this time period are displayed between the vertical dotted lines in Fig. \ref{fig:psp}, which overall spans a time interval of $3.5$ days. Specifically, magnetic field and plasma measurements were acquired with the fluxgate magnetometer of the FIELDS suite \citep{bale2016} and with the SPAN-Ai top-hat electrostatic analyzer of the Solar Wind Electrons Alphas \& Protons suite \citep[SWEAP,][]{kasper2016}, respectively, and then merged at $1$-minute resolution. As described by \citet{woodham2021}, the core proton velocity distributions have been fitted to a bi-Maxwellian function to obtain plasma moments (i.e. proton density, velocity and temperature).

The solar wind radial speed $V_{R}$ and the Alfv\'en speed $V_{A}$ are displayed in Fig. \ref{fig:psp}(a), showing that the plasma flow, despite being slow, is largely super-Alfv\'enic. Several spikes indicate bursts of accelerated wind. The magnetic field $\mathbf{B}$, in the inertial Radial Tangential Normal (RTN) frame, and its angles with the radial ($\theta_{RB}$) and solar wind ($\theta_{VB}$) directions are shown in Fig. \ref{fig:psp}(b) and (c), respectively. While the large-scale time profile of the magnetic field radial component is attributed to the expected $1/r^{2}$ scaling, with $r$ the radial distance displayed in Fig. \ref{fig:psp}(h), two sharp magnetic field $\sim180^{\circ}$ rotations clearly indicate crossings of the Heliospheric Current Sheet (HCS), so that during the interval under study the magnetic field is directed sunward. Several sharp rotations are present, coinciding with wind accelerations, although these do not reverse the field direction. In particular, there are no switchbacks \citep{bale2019} during the selected interval. The very small difference between $\theta_{RB}$ and $\theta_{VB}$ indicates that the solar wind velocity is nearly radial, thus supporting the assumption of a radially expanding coronal plasma underlying the ballistic assumption used in this work. Finally, Figs \ref{fig:psp}(d) -- (g) show the proton density $n_{p}$, temperature $T_{p}$, temperature anisotropy $T_{\perp,p}/T_{\parallel,p}$, and parallel plasma beta $\beta_{\parallel}=v_{th,\parallel}^{2}/V_{A}^2$ ($v_{th,\parallel}$ is the proton parallel thermal speed, where the suffix $p$ has been dropped for simplicity), all of which are relatively smooth during the selected interval (between the vertical dotted lines).

The temperature anisotropy $T_{\perp,p}/T_{\parallel,p}=1.72$ observed at PSP suggests a preferential plasma heating perpendicular to the magnetic field \citep[presumably through dissipation of low-frequency magnetohydrodynamic turbulence, see][for a comprehensive review of theoretical turbulence models and their convergence with PSP and SO observations]{zank2021}. At the observed parallel plasma beta $\beta_{\parallel}=0.66$ (i.e. the plasma is magnetically dominated), such an anisotropy is above the theoretical threshold for proton cyclotron instability \citep[see e.g.][]{hellinger2006,telloni2019}, which is therefore at work generating ion-cyclotron waves, as confirmed by windowed Fourier transforms of PSP magnetometer data in the $3-8$ Hz range (not shown).

\begin{figure*}[h]
	\begin{center}
		\includegraphics[width=\linewidth]{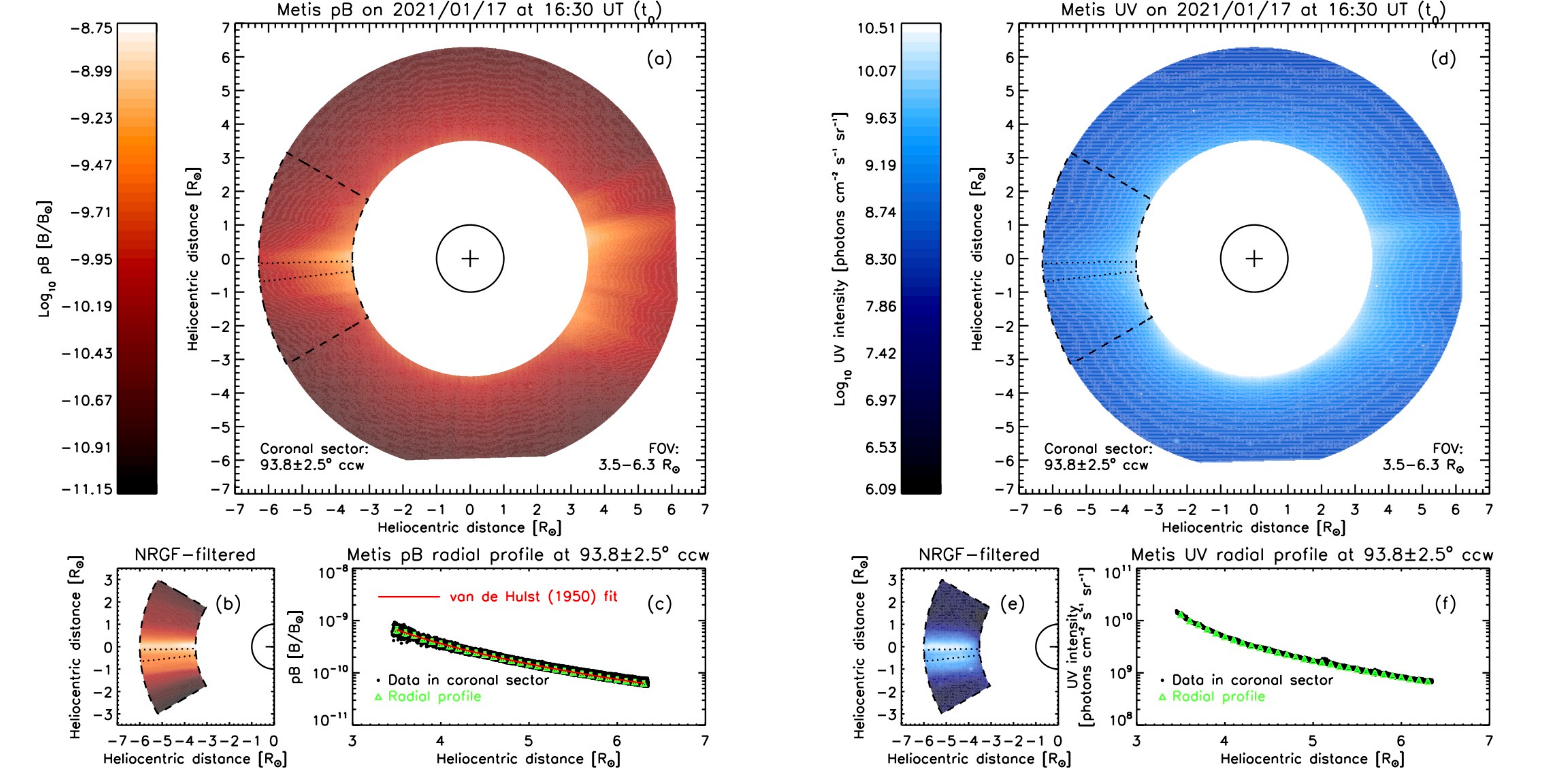}
	\end{center}
	\caption{Metis images of the pB (a) and UV \ion{H}{1} Ly$\alpha$ (b) emission of the solar corona observed on January $17$, $2021$ at $16:30$ UT ($t_{0}$) within the $3.5-6.3$ R$_{\odot}$ FOV. Data corresponding to the $93.8\pm2.5^{\circ}$ coronal sector in the PSP direction (delimited by dotted lines) are shown as black dots in panels (c) and (f) for pB and UV, respectively, where green triangles denote averaged values along the radial profile (corresponding standard deviation error bars are smaller than symbols size). A \citet{vandehulst1950} fit (red line), used to infer electron density, is applied to pB averages in panel (c). NRGF-processed pB and UV images in the angle sector ranging from $60^{\circ}$ to $120^{\circ}$ (indicated by dashed lines) are displayed in panels (b) and (e), respectively, to enhance coronal structures near the region under study.}
	\label{fig:metis}
\end{figure*}

The plasma stream detected with PSP originated from the equatorial region of the East limb solar corona imaged by Metis\footnote{The images were processed and calibrated according to the procedures adopted in \citet{romoli2021}.} on January $17$, $2021$ at $16:30$ UT simultaneously in polarized brightness (pB) and in \ion{H}{1} Ly$\alpha$ ultraviolet (UV) light (Fig. \ref{fig:metis}(a) and (d), respectively). This displays a typical solar minimum configuration with enhanced emission in the equatorial plane.


Two structures can be identified within the region of enhanced emission in both pB and UV. They are better distinguished by applying a Normalizing-Radial-Graded Filter \citep[NRGF,][]{morgan2006} to both images in an angle sector (delimitated by dashed lines) $\pm30^{\circ}$ wide centered on the equator (Fig. \ref{fig:metis}(b) and (e), respectively). The double equatorial structure is easily interpreted in light of the Wilcox Solar Observatory synoptic charts of the coronal magnetic field at the source surface\footnote{\href{http://wso.stanford.edu/synoptic/WSO-S.2240.gif}{http://wso.stanford.edu/synoptic/WSO-S.2240.gif}}. In the Metis POS the coronal current sheet was intersecting the equator. However, at about $\pm30^{\circ}$ in longitude from it were two small sheet warps, with maximum latitude extension to about $\pm10^{\circ}$, that might explain the double equatorial structure around the equator. This interpretation is also supported by the double HCS crossing experienced by PSP during its approach to the Sun (approximately on January $17$, $2021$ at $13:30$ UT and on January $19$, $2021$ at $21:00$ UT, see Fig. \ref{fig:psp}(c)).

The solar wind plasma crossing PSP came from the coronal sector $\pm2.5^{\circ}$ wide centered at $93.8^{\circ}$, which is the latitude reached by PSP on the POS of Metis during the observation time (Fig. \ref{fig:quadrature}(a)). This sector is delimited by dotted lines in Fig. \ref{fig:metis}(a), (b), (d), and (e). Black dots in Fig. \ref{fig:metis}(c) and (f) refer to pB and UV data in the selected coronal region, while green triangles denote the averaged values in steps of $0.1$ R$_{\odot}$, from $3.5$ to $6.3$ R$_{\odot}$. The pB and UV radial profiles are then used to infer the electron density $n_{e}$ and the radial outflow velocity $V_{R}$ of the coronal plasma which is later impinging on the PSP in-situ instruments.

\begin{figure*}[h]
	\begin{center}
		\includegraphics[width=\linewidth]{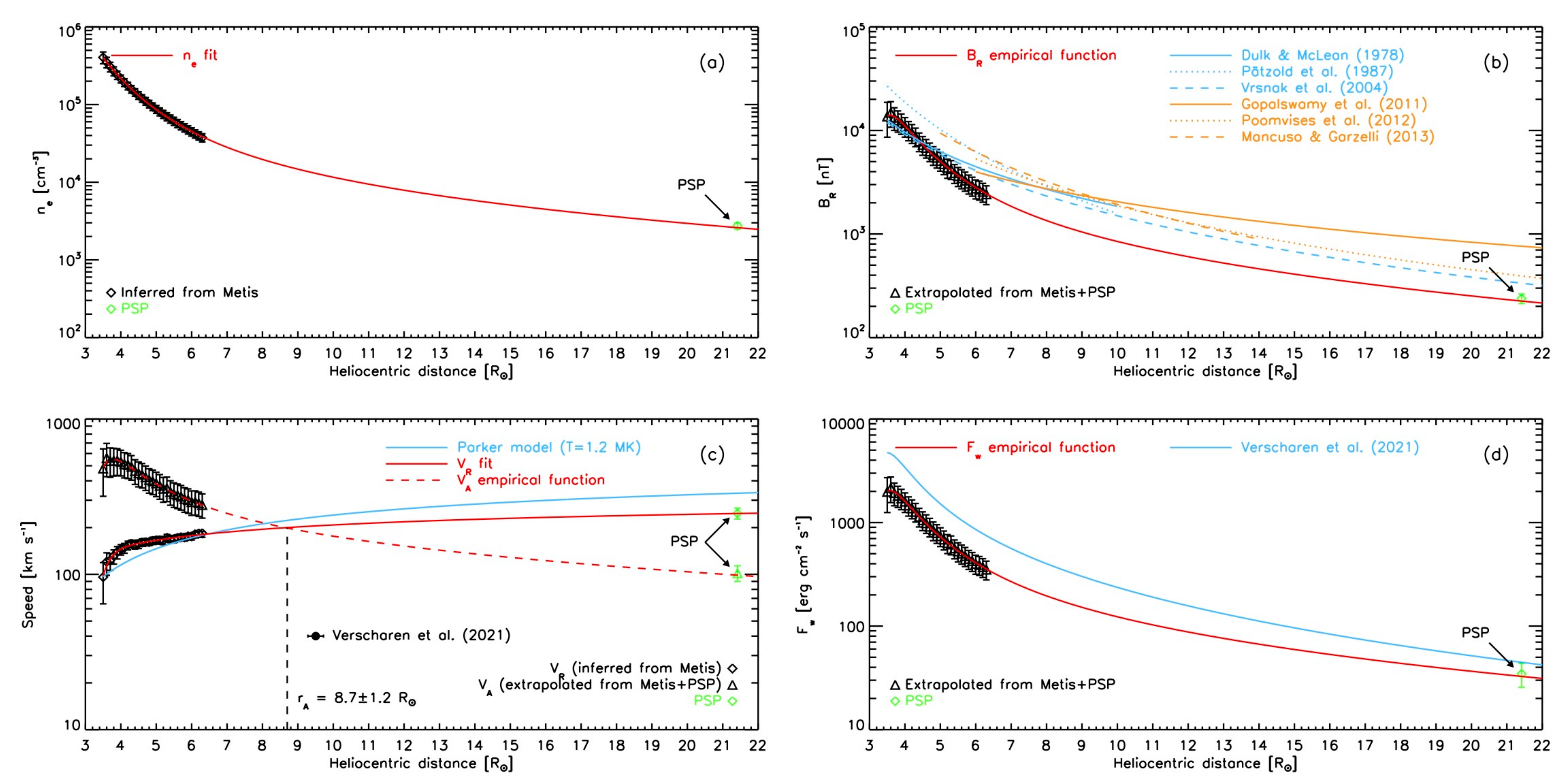}
	\end{center}
	\caption{Radial evolution of electron density $n_{e}$ (a), radial component of the magnetic field $B_{R}$ (b), radial coronal outflow $V_{R}$ and Alfv\'en speed $V_{A}$ (c), and bulk kinetic energy flux density of the solar wind $F_{w}$ (d), at coronal (black signs) and PSP (green sign, pointed by an arrow) heights. Diamonds denote direct Metis/PSP observations, while triangles mark extrapolated values based on conservation laws and PSP data as described in the text. Red solid or dashed curves denote corresponding fit and/or extrapolated empirical functions. The vertical dashed line and the blue solid curve in (c) indicate the inferred Alfv\'en radius $r_{A}=8.7\pm1.2$ R$_{\odot}$ \citep[compared with the result by][black full circle]{verscharen2021} and the Parker model for an isothermal corona with $T=1.2\times10^{6}$ K, respectively. As shown in the legend, empirical profiles previously reported in the literature are shown in (b) and (d) for comparison. Uncertainties of the derived quantities are drawn as error bars.}
	\label{fig:results}
\end{figure*}

Specifically, the electron density is derived by applying the inversion technique developed by \citet{vandehulst1950} and based on fitting pB data with a polynomial function (red line in Fig. \ref{fig:metis}(c)). These values are reported in Fig. \ref{fig:results}(a) (black diamonds) along with a polynomial fit (red line) extrapolated to the PSP location: the very good agreement with the PSP measurement \citep[green diamond: $n_{e,\rm{PSP}}=n_{p,\rm{PSP}}/0.95$, assuming a fully ionized plasma with $2.5$\% helium, according to][]{moses2020} corroborates the fact that the same element of plasma is indeed followed from the coronal region, where it is undergoing acceleration during its outward propagation, to the PSP position in the very inner heliosphere. The radial component of the coronal wind velocity on the POS is obtained by the Doppler dimming technique applied to a 3D solar coronal modeled on the basis of the so-inferred electron density and the measured UV \ion{H}{1} Ly$\alpha$ intensity \citep[the reader is referred to][for an exhaustive review on how to infer outflow velocities from UV radiative lines observed in the expanding corona and for the Metis first light observations of the coronal solar wind, respectively]{antonucci2020b,romoli2021}, under the following assumptions: (i) a hydrogen temperature $T_{p}=1.6\times10^{6}$ K \citep{antonucci2005}; (ii) a helium abundance of $2.5$\% \citep{moses2020}; (iii) a temperature anisotropy $T_{\perp,p}/T_{\parallel,p}=1.72$ as constrained by PSP observations; (iv) the electron temperature derived by \citet{gibson1999}; (v) the chromospheric \ion{H}{1} Ly$\alpha$ profile given by \citet{auchere2005}, scaled to the irradiance on January $17$, $2021$ ($5.42\times10^{15}$ photons cm$^{2}$ s$^{-1}$ sr$^{-1}$), as provided by the LASP Interactive Solar Irradiance Data Center\footnote{\href{https://lasp.corolaro.edu/lisird/data/composite\_lyman\_alpha/}{https://lasp.corolaro.edu/lisird/data/composite\_lyman\_alpha/}}. The resulting coronal $V_{R}$ values are displayed as black diamonds in Fig. \ref{fig:results}(c), where they are compared with the Parker outflow solution for an isothermal solar corona with $T=1.2\times10^{6}$ K \citep[][blue solid line]{parker1958}.

The coronal solar wind velocity undergoes a rapid increase with heliocentric distance, from $\sim80$ to $\sim150$ km s$^{-1}$, in the range from $3.5$ to $4.5$ R$_{\odot}$. Above this height, the acceleration slows down and a velocity of $\sim180$ km s$^{-1}$ is reached at $6.3$ R$_{\odot}$. Some residual acceleration persists up to the PSP location, where a solar wind speed of $247$ km s$^{-1}$ is measured locally (green diamond). Despite the overall good agreement with the purely hydrodynamic Parker model, in order to estimate as accurately as possible the transit time of the plasma volume from the Metis FOV ($r_{1}=3.5$ R$_{\odot}$) to the PSP location ($r_{2}=0.1$ au $=21.4$ R$_{\odot}$), Metis and PSP results were fitted to a parametric curve (solid red line: sum of a log-normal distribution and logarithmic quadratic function), which has no interpretative ambition, but aims only at defining an analytical profile. The transit time results in $t=\frac{1}{r_{2}-r_{1}}\int_{r_{1}}^{r_{2}}V(r)\rm{d}r=16.3\pm1.1$ h, thus confirming, as mentioned above, that PSP was measuring the same plasma observed with Metis during its propagation across the coronal region.


The simultaneous Metis and PSP observation of the same solar wind plasma is a noticeable step forward allowing the estimation of the coronal magnetic field beyond a few solar radii (not directly measurable in corona nowadays). This is achieved by applying the conservation of mass and magnetic flux (assuming flux-freezing) to Metis measurements of density and velocity (black diamonds in Fig. \ref{fig:results}(a) and (c)) and using the plasma and magnetic field data obtained at PSP, according to e.g. \citet{wang1995}. The so-extrapolated coronal magnetic field along the plasma flow $B_{R}$, in the equatorial region close to the coronal current sheet, is displayed in Fig. \ref{fig:results}(b) (black triangles): it varies from $\sim10^{4}$ to $\sim2\times10^{3}$ nT from $3.5$ to $6.3$ R$_{\odot}$. These values are somewhat lower than those previously reported in the literature \citep[][shown in Fig. \ref{fig:results}(b) with different colors and line styles]{dulk1978,patzold1987,vrsnak2004,gopalswamy2011,poomvises2012,mancuso2013} based solely on remote sensing observations (and thus suffering from LOS integration effects) and/or in-situ measurements very far from the Sun (thus not allowing a clear connection to the coronal source regions). In the present case, however, the accuracy of the values inferred for the coronal magnetic field is much higher due to the proximity of PSP, whose local measurements of the same plasma observed with Metis act as a constraint for extrapolations into the extended corona. Similarly, but considering $n_{e}$ and $V_{R}$ fits, an empirical function for $B_{R}$ (red line) can be extrapolated using the aforementioned conservation laws and PSP data.

Given the estimate of the coronal magnetic field, it is then also possible to extrapolate the Alfv\'en speed $V_{A}$ and, in turn, the Alfv\'en radius $r_{A}$ along the coronal current sheet, as done for $B_{R}$. Similarly, the solar-wind bulk kinetic energy flux density $F_{w}$ can be obtained considering the conservation of total energy along the stream line \citep[see e.g.][for details]{wang1995}. The corresponding empirical trends are also deduced. The extrapolated values for $V_{A}$ and $F_{w}$ are shown as black triangles in Fig. \ref{fig:results}(c) and (d), respectively. Specifically, $F_{w}$ varies from $\sim2\times10^{3}$ to $\sim4\times10^{2}$ erg cm$^{-2}$ s$^{-1}$ from $3.5$ to $6.3$ R$_{\odot}$ and is in fair agreement with the empirical profile deduced by \citet{verscharen2021} on the basis of a two-fluid magnetohydrodynamics framework and Ulysses observations, when neglecting enthalpy, magnetic stresses, and non-radial components of the solar-wind flow velocity (blue solid line in Fig. \ref{fig:results}(d)). Additionally, taking advantage of the $V_{A}$ empirical function (red dashed curve in Fig. \ref{fig:results}(c), deduced in the same way as that relative to $B_{R}$), the Alfv\'en radius $r_{A}$ can be estimated to be $8.7\pm1.2$ R$_{\odot}$ \citep[in agreement, within uncertainties, with the value of $9.5\pm0.2$ R$_{\odot}$ found by][]{verscharen2021}. This result is extremely interesting since, during the upcoming SO -- PSP quadrature on June $1$, $2022$, PSP will skim the Metis FOV that will extend from $5.6$ to $11.9$ R$_{\odot}$, thus likely allowing for the very first time the study of the transition of the coronal plasma from a sub-Alfv\'enic to a super-Alfv\'enic regime. Interestingly, the initial increase of $B_{R}$ (and, in turn, $V_{A}$ and $F_{w}$) up to about $3.8$ R$_{\odot}$ might depend on residual convergence of the magnetic field lines along the coronal current sheet before starting diverging \citep[see discussion in e.g.][]{schatten1969,pinto2017}.

It is worth reminding that the above estimates have been inferred assuming a helium abundance of $2.5$\% \citep{moses2020}. This value can be considered a lower limit (in fact, it refers to a particularly weak solar minimum). For the sake of completeness, it is therefore of interest to report how much the estimates of outflow velocity, magnetic field, kinetic energy flux, and Alfv\'en point vary assuming an upper limit of helium abundance of, say, $10$\%. The velocity and kinetic energy flux would be reduced on average by $5.7$\%, while the average reduction in the magnetic field and Alfv\'en speed would be $17.6$\% and $11.8$\%, respectively. Finally, the Alfv\'en radius would approach the Sun by $0.5$ R$_{\odot}$, i.e. it would be $8.2$ R$_{\odot}$.

This is the first time that crucial quantities such as the radial magnetic field in the outer corona, the Alfv\'en radius, and the bulk kinetic energy flux density of the solar wind have been extrapolated in the wind acceleration region with such an accuracy due to the opportunity offered by the first SO -- PSP quadrature. Moreover, it is worth noting that, as PSP approaches the solar corona more and more closely and finally enters it (crossing the Alfv\'en surface), the accuracy in deriving such quantities will further improve, providing a powerful tool to characterize, thanks to Metis observations, the corona in its expansion.

\acknowledgments
Solar Orbiter is a space mission of international collaboration between ESA and NASA, operated by ESA. D.T. was partially supported by the Italian Space Agency (ASI) under contract 2018-30-HH.0. L.S.V. was funded by the Swedish Contingency Agency, grant 2016-2102, and by SNSA, grant86/20. L.D.W. is supported by the STFC consolidated grant ST/S000364/1. G.P.Z. acknowledges the partial support of a NASA Parker Solar Probe contract SV4-84017, an NSF EPSCoR RII-Track-1 Cooperative Agreement OIA-1655280, and a NASA IMAP grant through SUB000313/80GSFC19C0027. O.P. was supported by the NASA grant 80NSSC 20K1829. L.S. was supported by a grant from the NASA Heliophysics Technology and Instrument Development for Science Program, NNH15ZDA001N-HTIDS, and also by the basic Research Funds of the Office of Naval Research. D.V. is supported by STFC Ernest Rutherford Fellowship ST/P003826/1 and STFC Consolidated Grant ST/S000240/1. The Metis program is supported by ASI under contracts to the National Institute for Astrophysics and industrial partners. Metis was built with hardware contributions from Germany (Bundesministerium f\"ur Wirtschaft und Energie through the Deutsches Zentrum f\"ur Luft- und Raumfahrt e.V.), the Czech Republic (PRODEX) and ESA. The FIELDS and SWEAP teams acknowledge support from NASA contract NNN06AA01C. Parker Solar Probe data was downloaded from the NASA's Space Physics Data Facility (\href{https://spdf.gsfc.nasa.gov}{https://spdf.gsfc.nasa.gov}). D.T. wishes to thank Masaru Nakanotani for his help in sketching cartoon of Fig. \ref{fig:quadrature}.
\par

\end{document}